\documentclass[%
 reprint,
 superscriptaddress,
 nofootinbib,
 amsmath, amssymb,
 prl,
]{revtex4-2}

\usepackage{graphicx}
\usepackage{dcolumn}
\usepackage{bm}
\usepackage{hyperref}

\usepackage{siunitx}
\usepackage{subcaption}

\begin{document}

\title{Pulsed Electron Lenses for Space Charge Mitigation}

\author{Adrian Oeftiger}
 \email{a.oeftiger@gsi.de}
\affiliation{GSI Helmholtzzentrum f\"ur Schwerionenforschung GmbH, Planckstrasse 1, 64291 Darmstadt, Germany}
\author{Oliver Boine-Frankenheim}
\affiliation{GSI Helmholtzzentrum f\"ur Schwerionenforschung GmbH, Planckstrasse 1, 64291 Darmstadt, Germany}
\affiliation{Technische Universit\"at Darmstadt, Schlossgartenstrasse 8, 64289 Darmstadt, Germany}

\date{\today}

\begin{abstract}
  To produce ultimate high-brightness hadron beams, synchrotrons need to overcome a most prominent intensity limitation i.e., space charge. This Letter characterizes the potential of pulsed electron lenses in detailed 3D tracking simulations, key to which is a realistic machine and space charge model. The space charge limit, imparted by betatron resonances, is shown to be increased by up to 50\% using a low symmetric number of electron lenses in application to the FAIR SIS100 synchrotron. Conceptually, a 100\% increase is demonstrated with a larger number of electron lenses, which is found to rapidly saturate near the theoretical 2D limit.
\end{abstract}

\maketitle

Today, synchrotrons are the preferred tool to produce focused ion beams of highest intensities with kinetic energies starting from the GeV range.
Their performance is ultimately limited by space charge effects at low (injection) beam energies \cite{handbook}. In particular, periodic resonance crossing is induced by space charge detuning coupled with synchrotron motion and adversely affects beam quality over typical storage times  \cite{PhysRevSTAB.13.114203,Bartosik_2020,PhysRevAccelBeams.23.091001,PhysRevAccelBeams.24.044001,PhysRevAccelBeams.25.054402}. This limitation necessitates mitigation to increase the performance of future machines and upgrades, as discussed in the review article by \textcite{wei}.
Electron lenses, employed as insertion devices within hadron synchrotrons, have effectively addressed issues such as beam halo cleaning \cite{beamhalocleaning} and beam-beam compensation \cite{PhysRevLett.99.244801,beambeamlenses}, see Refs.~\cite{shiltsev2015electron,Shiltsev_2021} for a comprehensive overview.
For the nonlinear beam-beam compensation, the transverse electron beam profile is Gaussian to match the transverse ion beam profile. This idea was soon suggested to compensate for the bunch self-fields instead of a colliding second beam \cite{burov2001}.
However, the continuous nature of space charge in contrast to localized beam-beam lensing makes the compensation more challenging. 
Placing one or even several such transversely nonlinear electron beam elements in every basic focusing cell, optimum configurations for strong space charge compensation have been identified in several studies \cite{Alexahin_2021,Stern_2021}. A realistic scenario should ideally provide compensation with only a few electron lenses for the entire synchrotron though. First simulation studies in 2D approximation concluded that---unless an electron lens is placed in every focusing cell---transversely nonlinear electron lenses typically drive prohibitively strong systematic nonlinear resonances  \cite{BOINEFRANKENHEIM2018122}. 
To avoid these, the authors postulated that electron beams with a transversely homogeneous distribution (which thus exert a linear beam-beam force) provide the most efficient space charge mitigation: longitudinal modulation of the electron beam then tackles the problem of space-charge induced periodic resonance crossing by suppressing the variation of the space charge strength along the longitudinal bunch profile. 
A theoretical asymptotic limit to this scheme is given by 2D resonance dynamics of a longitudinally ``frozen'' bunch without synchrotron motion.
This Letter demonstrates the efficacy of such pulsed linear electron lenses: for the first time, the space charge limit is characterized depending on the electron lens configuration in a realistic simulation scenario. 
On the basis of a comprehensive model of a space-charge limited synchrotron, we quantify the extent to which pulsed linear electron lenses increase the maximum achievable intensity.

Commonly employed approaches to mitigate space charge effects include longitudinal bunch shape flattening (to reduce tune spread) and resonance compensation (to reduce resonance stopband widths).
Bunch shaping techniques based on dual-harmonic rf systems \cite{garoby1997longitudinal} and hollow phase-space distributions \cite{Oeftiger:2242787} increase the space charge limit by approximately $20\%$ to $30\%$ \cite{PhysRevAccelBeams.25.054402}.
Similarly, resonance compensation typically achieves results of comparable magnitude \cite{fedotov_rescomp,Hernandez_2018,jparc-rescomp,RABUSOV2022167290}. 

In general, the self-fields of a bunch in a synchrotron counteract externally applied transverse focusing, reducing the transverse oscillation frequency of particles in the bunch, also known as the incoherent betatron tune.
For a 2D homogeneous particle distribution in the transverse plane (the ``K-V distribution''), the space charge field is linear and all particles experience the same tune shift
\begin{equation} \label{eq: sc tune shift}
    \Delta Q_{x,y}^\mathrm{KV} = - \frac{r_{\mathrm{c}} \lambda}{2\beta_0^2\gamma_0^3} \oint \frac{ds}{2\pi} \frac{\beta_{x,y}(s)}{\sigma_{x,y}(s) \bigl(\sigma_x(s)+\sigma_y(s)\bigr)} \quad ,
\end{equation}
where $r_{\mathrm{c}}$ denotes the classical particle radius, $\lambda$ the line density (number of particles per longitudinal unit length), $\beta_0$ the ion beam speed in units of the speed of light $c$, $\gamma_0$ the corresponding Lorentz factor, $\beta_{x,y}(s)$ the horizontal or vertical betatron function along the path length $s$ around the accelerator, and $\sigma_{x,y}$ the local horizontal or vertical rms beam size.
\textcite{kv} first derived Eq.~\eqref{eq: sc tune shift} assuming smooth focusing, i.e., no dependency on $s$; the expression here provides more accurate results for alternate-gradient focusing in synchrotrons \cite{oeftigerthesis} as required for the quantitative analysis below.

The space charge limit is a consequence of the space-charge induced tune spread, which increases with the bunch intensity $N$. Above a certain intensity, the extending beam response reaches nearby located resonant tunes driven by magnetic field imperfections.
As a consequence, beam quality can degrade in terms of a growing transverse emittance and particle loss in the physical machine aperture. 

For a 3D bunch distribution affected by space charge, two aspects cause a spread of incoherent betatron tunes.
The first reason lies in the nonlinearity of the transverse self-fields in the ion bunch, in the case of a non-homogeneous transverse distribution. 
Synchrotrons typically produce nearly Gaussian distributed ion bunches, for which the maximum extent of the space-charge tune spread reaches twice the linear tune shift, $\left|\Delta Q^{SC}_{x,y}\right| = 2\cdot \left| \Delta Q^{KV}_{x,y}\right|$.
Particles at the 3D bunch center are subject to maximum detuning.

The second reason is the longitudinal bunch shape, along which the line density $\lambda$ varies, $0\leq\lambda(z)\leq\lambda_\mathrm{max}$. 
A Gaussian shaped bunch of rms bunch length $\sigma_z$ features a maximum line density of $\lambda_\mathrm{max} = N/(\sqrt{2\pi}\sigma_z)$.

Pulsed electron lenses aim to reduce the space-charge tune spread by suppressing the longitudinal variation of $\lambda(z)$ \cite{osti_1193236,BOINEFRANKENHEIM2018122,Artikova_2021,Schulte-Urlichs_2023}: a co-propagating electron beam pulse is shaped to longitudinally match the circulating hadron bunch. 
In the particular case of pulsed \emph{linear} electron lenses, as employed in our study, the electron beam is distributed homogeneously in the transverse plane.

\begin{figure}[t]
    \centering
    \includegraphics[width=\linewidth]{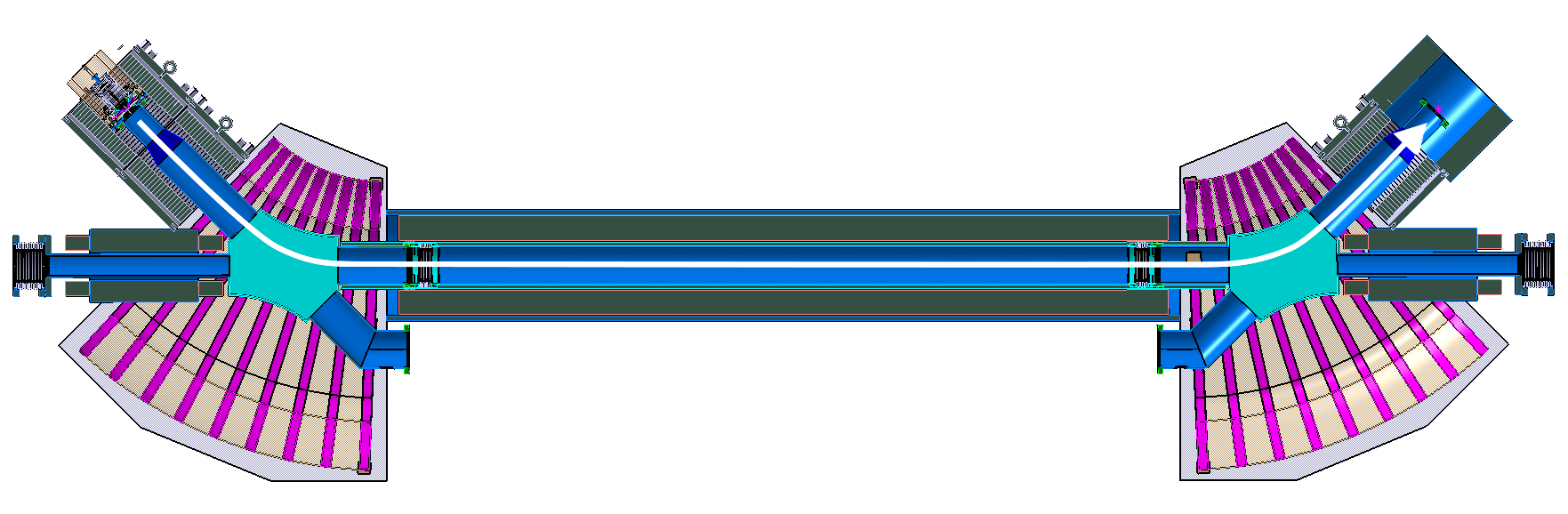}
    \caption{Layout of the GSI SIS18 pulsed electron lens (image from Ref.~\cite{Schulte-Urlichs_2023}).
    The ion beam traverses the electron lens insertion from the left to the right in the straight beam pipe. 
    The electron beam is generated in the cathode on the top left and a modulation grid shapes the electron pulse, which then follows the white arrow: a toroid (violet) guides into the $L=\SI{3.36}{\meter}$ long interaction region with focusing solenoid fields before a second toroid directs the electron beam away from the circulating ion bunch onto the collector on the top right.}
    \label{fig:layout}
\end{figure}

A 3D technical layout illustrating such a space-charge compensation electron lens device is depicted in Fig.~\ref{fig:layout}, developed here for the SIS18 at GSI \cite{Schulte-Urlichs_2023}.
The hardware closely resembles an electron cooler device, with the major difference to operate at shifted electron beam speed $\beta_e\neq\beta_0$ to prevent cooling. The SIS18 demonstrator provides an electron beam with a \SI{10}{\ampere} peak current at \SI{30}{\kilo\electronvolt} kinetic energy.
Typically, hadron bunches in synchrotrons significantly exceed the length of the electron lens interaction region $L$ and typically $\beta_{x,y}\gg L$.
Also, the weak local coupling and focusing effects from the electron lens guiding magnets can be neglected to first approximation.
It follows that the electron lens effect on the hadron beam can be modeled as a thin time-modulated linear kick.

The maximum beam-beam tune shift induced on the ion beam by a set of $n_\mathrm{el}$ e-lenses at locations $s_k$ is positive and reads \cite{burov2001}
\begin{equation}
    \Delta Q^\mathrm{e}_{x,y} = \frac{1}{4\pi}\sum\limits_{k=1}^{n_\mathrm{el}} \beta_{x,y}(s_k) \frac{r_{\mathrm{c}}}{Ze} \frac{I_e}{\sigma_e^2 \gamma_0} \frac{1-\beta_e\beta_0}{\beta_e} \frac{L}{\beta_0 c}\quad ,
\end{equation}
with $I_e$ the e-lens current, $Z$ the ion charge number and $e$ the elementary charge.
This requires, firstly, the radii $\sigma_e$ of the (round and transversely homogeneous) electron beam to cover the ion beam (up to $2\sigma_{x,y}$) and, secondly, shaping the electron pulse to match the profile of the ion bunch $\lambda(z)$. 
As external elements, the e-lenses also shift the tune of the ion beam's dipole moment upward by $\Delta Q^\mathrm{e}_{x,y}$. Direct space charge, in contrast, does not affect the tune of the dipole moment, since the inter-particle Coulomb forces always sum up to zero. However, potential indirect space charge contributions via the surrounding environment may reduce the dipole tune and, consequently, diminish this e-lens effect.

To parameterize the strength of the e-lens configuration, the linear compensation degree $\alpha$ is defined as
\begin{equation}
    \alpha=\frac{\Delta Q^\mathrm{e}_y}{|\Delta Q^\mathrm{KV}_y|}
\end{equation}
where $\alpha=1$ indicates a full compensation of the linear space charge tune shift $\Delta Q^\mathrm{KV}_y$ by linear e-lenses.
Reference~\cite{BOINEFRANKENHEIM2018122} established a theoretical optimum at $\alpha=0.5$ for compensation purposes.

\begin{figure}[t]
    \centering
    \includegraphics[width=0.95\linewidth]{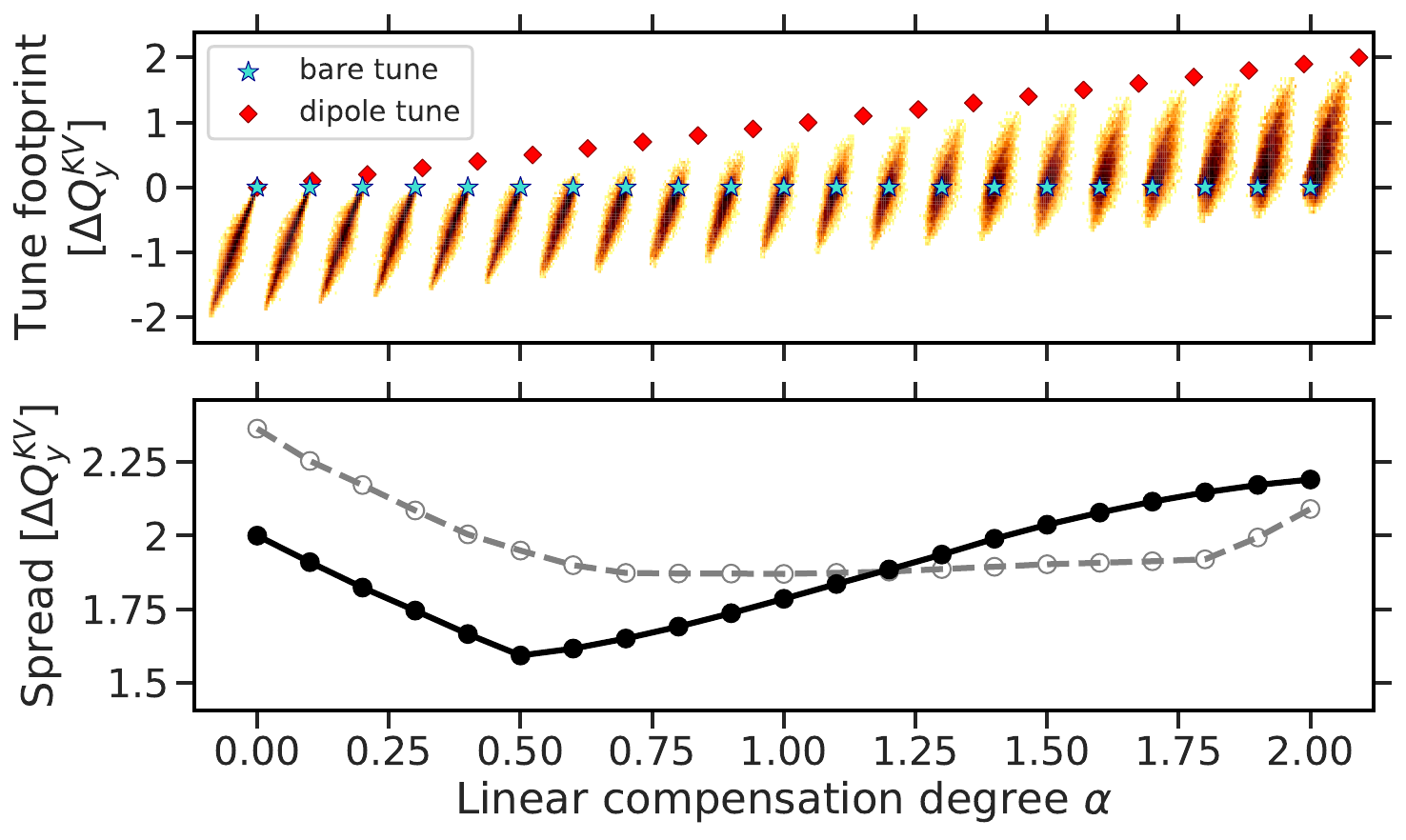}
    \caption{Top: incoherent space charge tune footprints against electron lens strength. Bottom: corresponding vertical extent of footprint (black solid line: only space charge, grey dashed line: including chromaticity). }
    \label{fig: tunespreads elens}
\end{figure}

\begin{figure*}[tp]
    \begin{subfigure}[T]{0.24\linewidth}
        \centering
        \includegraphics[width=\linewidth]{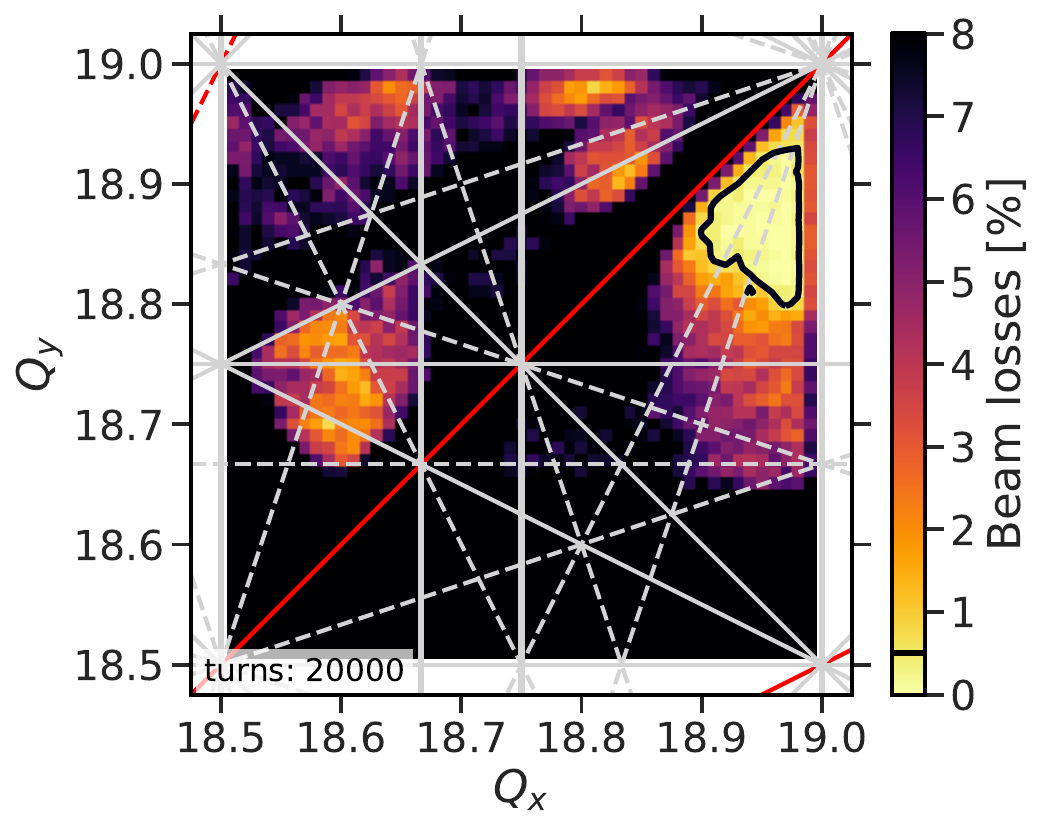}
        \caption{
        $\alpha=0$}
        \label{fig: ffsc no elens}
    \end{subfigure}
    \hfill
    \begin{subfigure}[T]{0.24\linewidth}
        \centering
        \includegraphics[width=\linewidth]{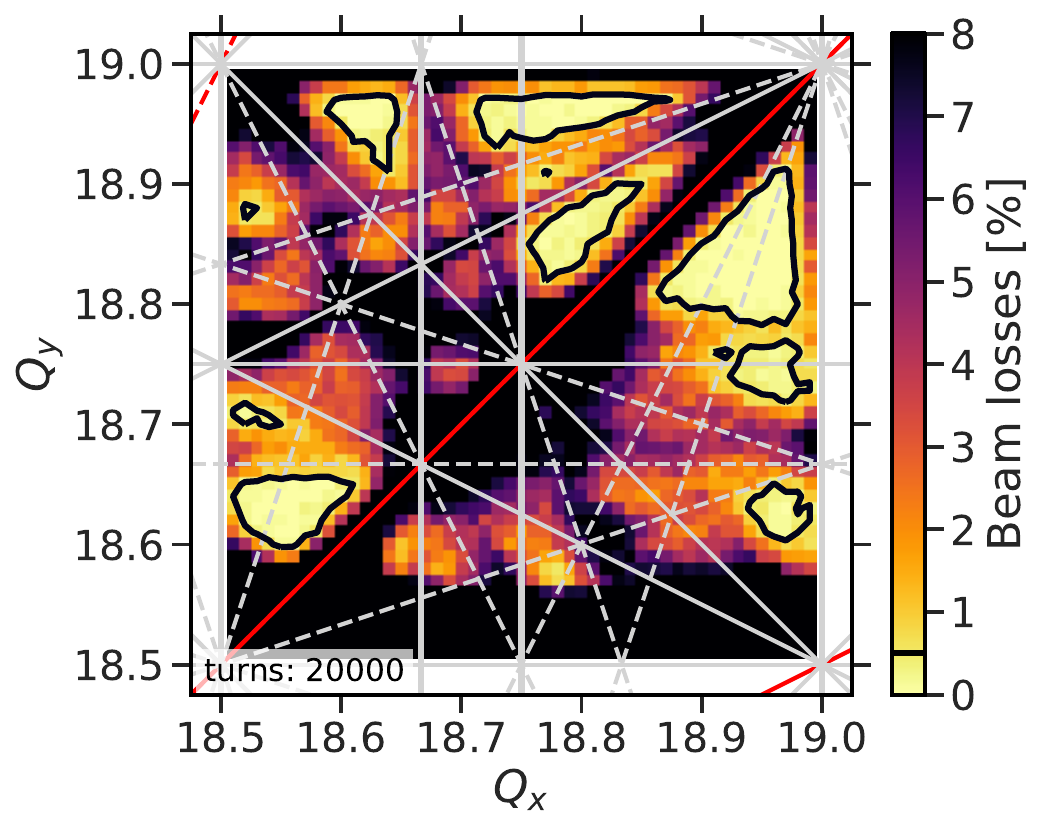}
        \caption{
        $\alpha=0.5$}
        \label{fig: ffsc alpha 0p5}
    \end{subfigure}
    \hfill
    \begin{subfigure}[T]{0.24\linewidth}
        \centering
        \includegraphics[width=\linewidth]{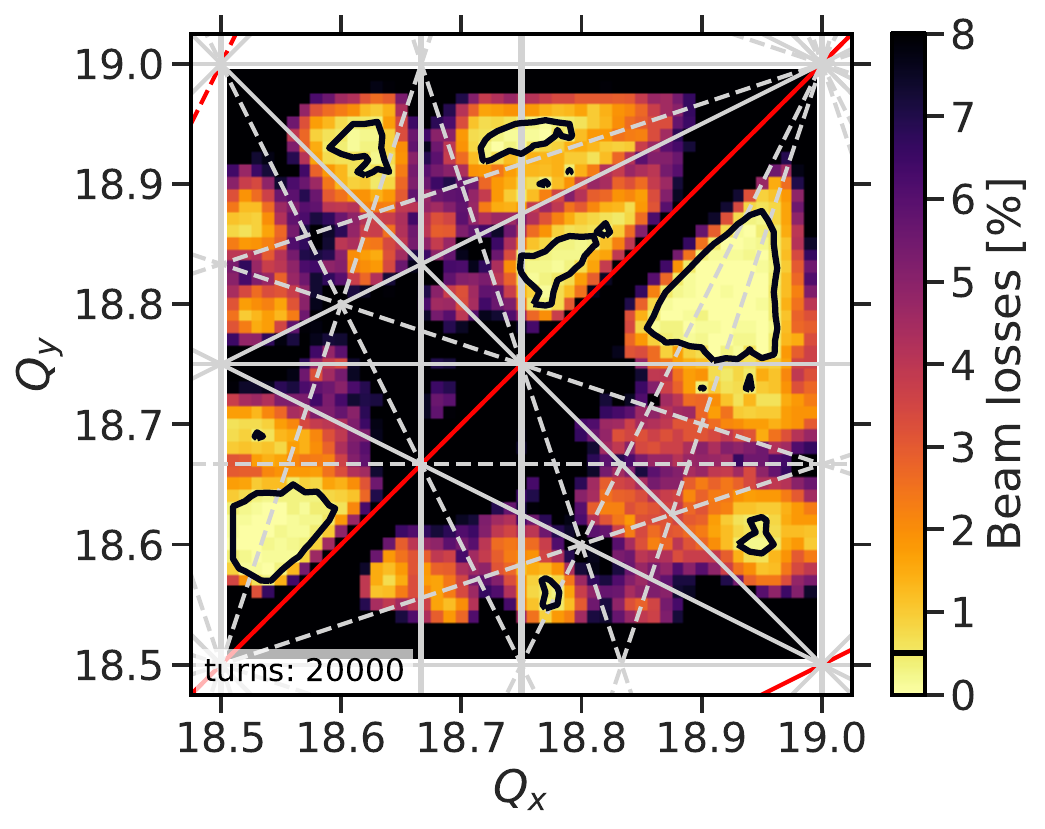}
        \caption{
        $\alpha=0.7$}
        \label{fig: ffsc alpha 0p7}
    \end{subfigure}
    \hfill
    \begin{subfigure}[T]{0.24\linewidth}
        \centering
        \includegraphics[width=\linewidth]{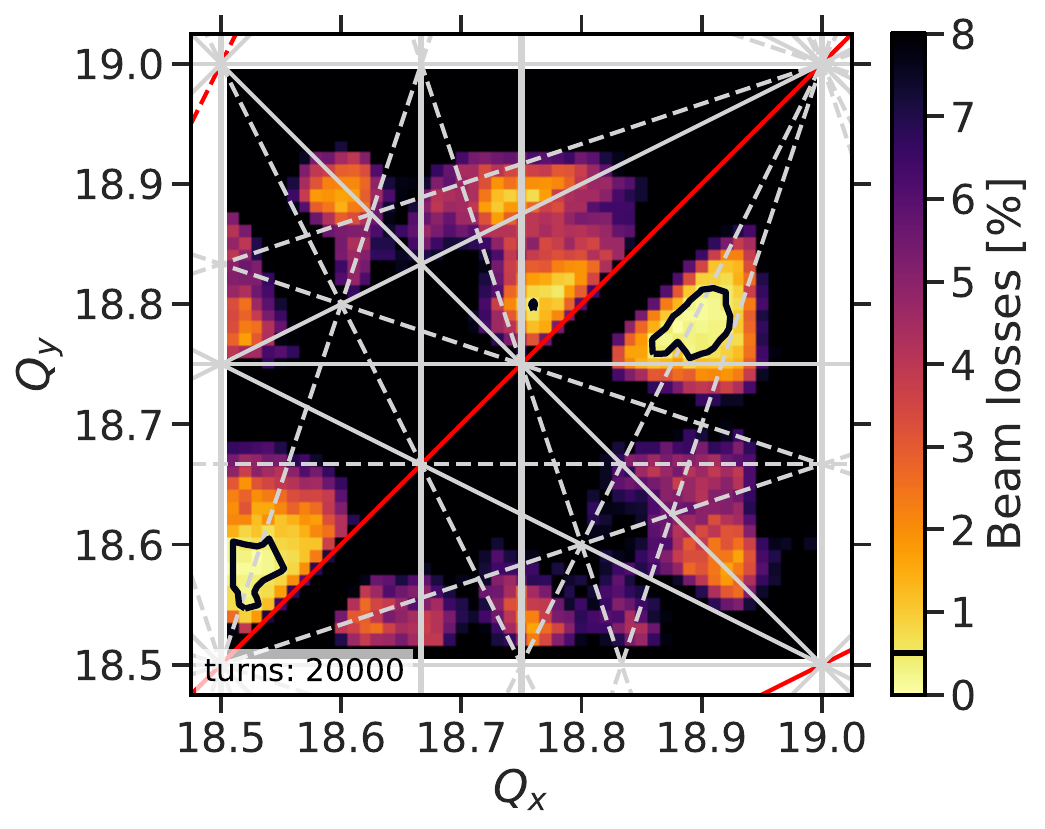}
        \caption{
        $\alpha=1$}
        \label{fig: ffsc alpha 1}
    \end{subfigure}
    \caption{Beam loss in transverse tune space for SIS100 with sixfold superperiodicity and 
    three electron lenses, the plotted tune diagrams show four different linear compensation degrees $\alpha$ at FAIR design bunch intensity $N=N_0$.}
    \label{fig: ffsc alpha}
\end{figure*}

Figure \ref{fig: tunespreads elens} illustrates the influence of pulsed linear e-lenses on the tune footprint of a Gaussian bunch as a function of the degree of linear compensation $\alpha$. The upper panel presents a 2D histogram depicting the simulated ion bunch tune footprint (cf.\ the Appendix). The footprint is shown relative to the linear machine tune without the electron lens effect (indicated by a blue star), termed the ``bare'' machine tune. Darker colors signify higher density tune areas. A sequence of tune footprints is displayed for increasing $\alpha$. The correspondingly increasing dipole tunes of the ion bunch are marked with red diamonds. 
At $\alpha=0$, the tune footprint of the 3D Gaussian bunch is solely provided by the defocusing effect of the space charge and vertically spans $-2\Delta Q^\mathrm{KV}_y$, as expected. Conversely, at $\alpha=2$, the entire tune footprint extends above the bare machine tune. Here, the particles at the bunch center, experiencing the maximum space charge tune shift, are fully compensated back to the bare machine tune. 
Notably, the overall tune footprint is located below the dipole tune for any $\alpha$.

The lower panel of Fig.~\ref{fig: tunespreads elens} presents the vertical footprint extension, i.e., the spread in units of the linear KV tune shift according to Eq.~\eqref{eq: sc tune shift}. The solid black line corresponds to the upper panel footprints and exhibits a minimum around $\alpha=0.5$, confirming the theoretical estimate from Ref.~\cite{BOINEFRANKENHEIM2018122}. In practice, particle tunes will also be influenced by longitudinal momentum deviation $\delta=(p-p_0)/p_0$ (the lattice ``chromaticity'' is defined to first order by $Q'_{x,y}=dQ_{x,y}/d\delta$) and lattice nonlinearities, alongside space charge and the e-lenses. The grey dotted line incorporates typical natural chromatic detuning and is referenced later in this Letter. Remarkably, including chromaticity results in a plateau of minimal tune spread between $0.6\lesssim \alpha \lesssim 1.8$ rather than a single minimum. 

To demonstrate the mitigation concept, we consider, as an example, the heavy-ion synchrotron SIS100 \cite{Spiller:2014qqa} currently under construction at the Facility for Antiproton and Ion Research (FAIR). SIS100 features $S=6$ arcs and straight sections. The expected space charge limit of SIS100 has been determined through an extensive 3D tracking simulation study \cite{PhysRevAccelBeams.25.054402}. 
A key aspect is to correctly model the beam response to error resonances by means of a realistic magnetic field error model, which in the case of SIS100 is based on cold bench measurements of the main magnets. We employ identical beam parameters, machine configuration, field error distribution, and simulation tool set to investigate the potential of space charge mitigation via pulsed linear e-lenses, as summarized in Table~\ref{tab: params}. Space charge is modeled using the fixed frozen 3D Gaussian field map following the approach established in Ref.~\cite{PhysRevAccelBeams.25.054402}. For this choice, the location and extent of (incoherent) resonance stopbands are accurately modeled, although exact figures in case of significant beam loss are not meaningful. We assume an electron beam in the lenses with the same (Gaussian) pulse shape as the ion bunch. To identify whether a working point is affected by a resonance stopband, a 3D Gaussian distributed bunch is tracked for 20000 turns, corresponding to an eighth of the foreseen \SI{1}{\sec} SIS100 heavy-ion accumulation plateau. Due to tight aperture constraints in the SIS100 (at approximately three rms ${}^{238}$U$^{28+}$ beam sizes), significant emittance growth immediately translates into beam loss. Therefore, a finite beam loss figure suffices as an observable to identify resonance stopbands. 

\begin{table}[t]
    \centering
    \caption{Simulation parameters for SIS100 ion beams.}
    \renewcommand{\arraystretch}{1.1}
    \label{tab: params}
    \begin{tabular}{lc}
        \hline\hline
        Parameter & Value \\ \hline
        FAIR design ${}^{238}$U${}^{28+}$ bunch intensity $N_0$ & $\SI{6.25d10}{}$ \\
        Max.\  $\Delta Q^\mathrm{SC}_{x,y}$ (at $N_0$) & $(-0.21,-0.30)$ \\
        rms ion and electron bunch length $\sigma_z$ & \SI{13.2}{\meter} \\
        rms chromatic tune spread $Q'_{x,y}\cdot\sigma_{\Delta p/p_0}$ & $0.01$ \\
        Synchrotron tune $Q_\mathrm{s}$ & \SI{4.5d-3}{} \\
        Circumference $C$ & \SI{1083.6}{m} \\
        Number of basic focusing cells & 84 \\
        Superperiodicity (arcs) $S$ & 6  \\
        Beam rigidity $B\rho$ & \SI{18.2}{\tesla\meter} \\
        Relativistic $\beta_0$ factor & 0.568 \\
        \hline\hline
    \end{tabular}
\end{table}

First, we present the impact of the compensation degree $\alpha$ on beam losses due to betatron resonances in the tune space. Simulations were conducted by scanning the relevant tune quadrant in $0.01$ tune steps at FAIR design heavy-ion bunch intensity $N_0$. Figure~\ref{fig: ffsc alpha} displays the beam loss results plotted against the horizontal and vertical bare machine tunes $Q_x$ and $Q_y$. Dark areas signify high beam loss, implying the presence of resonance stopbands, while yellow indicates good working point areas with low beam loss. Working points within the black contours exhibit less than $0.5\%$ beam loss and are practically unaffected by resonances. 

Figure~\ref{fig: ffsc no elens} depicts the reference scenario without space charge compensation, as discussed in Ref.~\cite{PhysRevAccelBeams.25.054402}. For Figs.~\ref{fig: ffsc alpha 0p5}, \ref{fig: ffsc alpha 0p7} and \ref{fig: ffsc alpha 1}, three e-lenses are placed at symmetric locations in every second straight section and powered at half, $70\%$, and full linear space charge compensation degree, respectively. 
SIS100 is set to operate at natural chromaticity at injection, and the overall tune spread is approximately consistent for all three compensation cases. The grey line in Fig.~\ref{fig: tunespreads elens} corresponds precisely to this simulation scenario. 
The rationale behind why $\alpha=0.5$ yields a larger low-loss area than $\alpha=1$ becomes evident when observing that the location of stopbands scales differently with $\alpha$ depending on the resonance type: for instance, compare the black Montague stopband around $2Q_y-2Q_x=0$ (shifting downward with $\alpha$) and the horizontal half-integer stopband around $2Q_x=38$ (entering from the right with increasing $\alpha$). 
We infer that the optimal compensation degree $\alpha$ depends on the specific resonances limiting the high-intensity working point region and, consequently, generally on the actual machine and beam parameters.

Next, we shift the focus to the space charge limit for various configurations of e-lenses. Pulsed linear e-lenses act as localized focusing errors and contribute to envelope beating (equivalent to $\beta$-beating), as discussed in Ref.~\cite{BOINEFRANKENHEIM2018122}. A configuration of $n_\mathrm{el}\leq S$ symmetrically placed e-lenses drives the half-integer resonances
\begin{equation}
    2Q_{x,y}=n_\mathrm{el}\cdot m \quad \text{for}~m\in\mathbb{Z} \quad .
\end{equation}

A single or two e-lenses result in excessively high beam losses across the tune quadrants with $\alpha=0.5$ at $N_0$ in the SIS100 scenario, effectively reducing the space charge limit. Three lenses drive $Q_{x,y}=18$ and $Q_{x,y}=19.5$, both sufficiently distant from the designated fast-extraction tune quadrant $18.5\leq Q_{x,y}\leq 19$. Thus, $n_\mathrm{el}=3$ represents the minimum useful number of e-lenses. For a stable working point outside the half-integer stopbands, the maximum vertical envelope beating $\max\limits_{s}[(\sigma^2_{y,n_\mathrm{el}}-\sigma^2_{y,n_\mathrm{el}=0})/\sigma^2_{y,n_\mathrm{el}=0}]$ reaches a significant $62\%$ for one e-lens and $21\%$ for two e-lenses, but remains below $10\%$ for three and more e-lenses. The $\beta$-beating analysis in Ref.~\cite{BOINEFRANKENHEIM2018122} similarly concludes that space charge compensation in high-intensity synchrotrons requires $n_\mathrm{el}\gtrsim 3$.

\begin{figure}[b]
    \centering
    \includegraphics[width=0.8\linewidth]{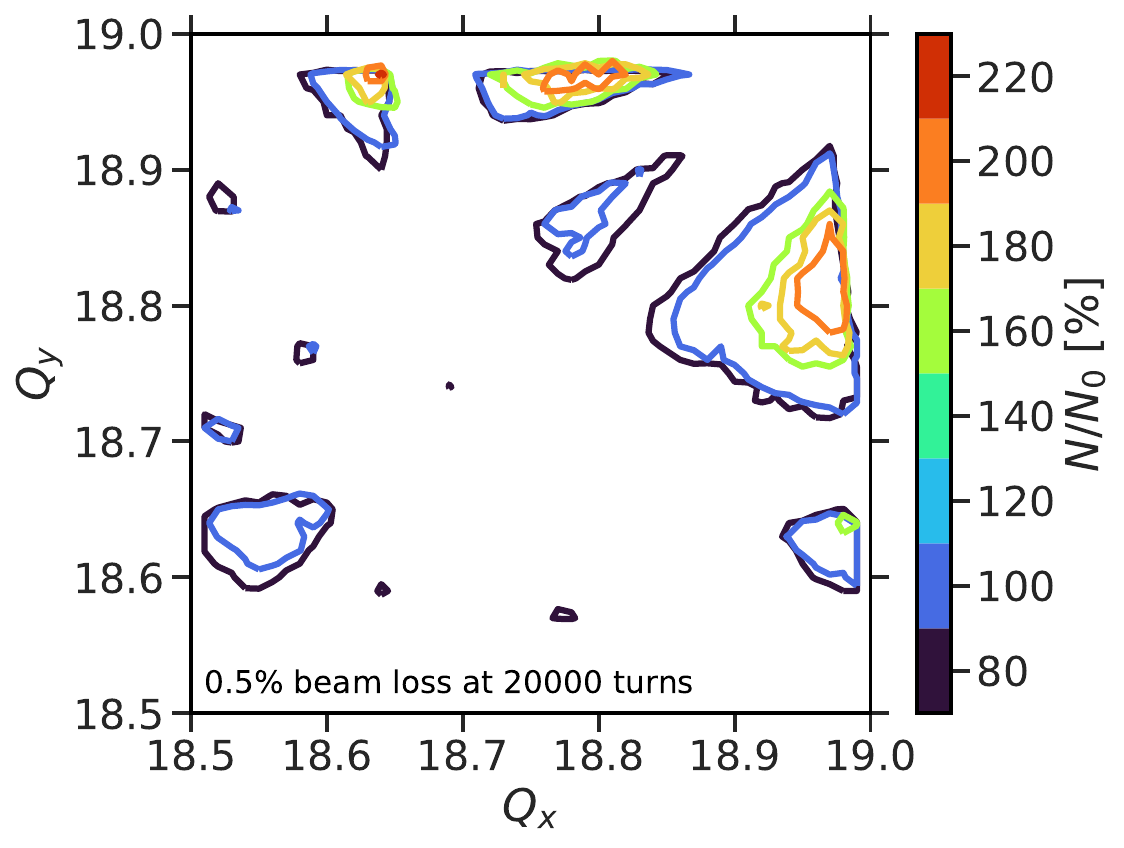}
    \caption{Low-loss tune area contours at various intensities for six electron lenses ($\alpha=0.5$).}
    \label{fig: sclimit 6el}
\end{figure}

To assess the space charge limit of an e-lens configuration, beam loss simulations are conducted analogously to Fig.~\ref{fig: ffsc alpha} across the tune quadrant $18.5\leq Q_{x,y}\leq 19$. These scans are performed for increasing bunch intensity $N$ while keeping all other parameters constant. Therefore, space charge increases linearly as $\Delta Q^\mathrm{SC}_{x,y}\propto N$. Figure~\ref{fig: sclimit 6el} illustrates the results for $n_\mathrm{el}=6$ e-lenses, displaying contours of low-loss tune areas where beam loss remains below $0.5\%$, with color encoding intensity. 
These low-loss areas shrink with increasing intensity, virtually vanishing up to a single working point at $N=2.2\cdot N_0$. 
From an operational perspective, it appears reasonable to define the space charge limit as the intensity above which the largest low-loss area diminishes below a certain small threshold size, e.g., a lower bound of 10 working points. Hence, for the $n_\mathrm{el}=6$ configuration, the space charge limit is reached at $N=2.1\cdot N_0$.

\begin{figure}[tp]
    \centering
    \includegraphics[width=0.8\linewidth]{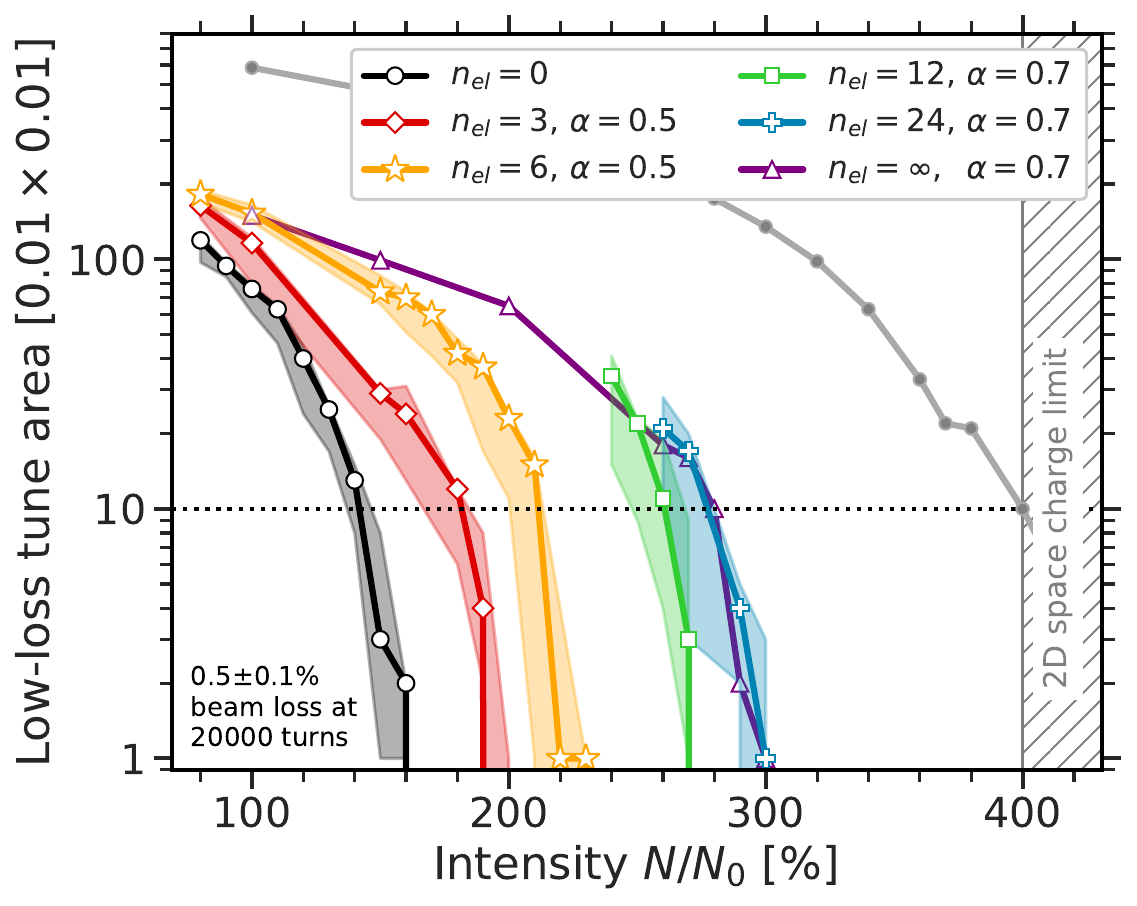}
    \caption{SIS100 space charge limit. Low-loss tune area against intensity for various electron lens configurations.}
    \label{fig: sclimit}
\end{figure}

This procedure is repeated for various e-lens configurations, with $n_\mathrm{el}=12$ and $24$ representing two and four e-lenses per straight section, respectively. For each intensity per configuration, the size of the largest low-loss tune area is evaluated by counting the corresponding working points. Figure~\ref{fig: sclimit} presents the key results of this Letter: solid lines plot the sizes of said low-loss tune area against the bunch intensity for the SIS100, where each colored line represents a particular e-lens configuration. The markers correspond to the low-loss tune area size for the threshold of $0.5\%$ beam loss. 
Table \ref{tab: results} summarizes the corresponding identified space charge limits for each e-lens configuration, where the markers remain above the lower bound of 10 working points indicated by the dotted line.
Notably, the space charge limit steadily increases with the number of e-lenses starting from the black $n_\mathrm{el}=0$ case without e-lenses up to the blue $n_\mathrm{el}=24$ case, irrespective of the chosen lower bound on the ordinate. This observation remains consistent regardless of the beam loss threshold figure: a shaded area of the same color is plotted around the markers, connecting the intervals of low-loss area sizes between $0.4\%$ and $0.6\%$ beam loss threshold per intensity.
The findings are validated by self-consistent particle-in-cell (PIC) simulations, cf.\ the Appendix.

At this point, several observations and comparisons are due. Regarding the optimal linear compensation degree $\alpha$, for $n_\mathrm{el}=3$ and $6$ e-lenses the compensation with $\alpha=0.5$ yields the largest tune area, while for the (more academic) cases of $n_\mathrm{el}=12$ and $24$, a maximum space charge limit could be achieved with $\alpha=0.7$.
To explore the \emph{continuous} e-lens limit, the purple $n_\mathrm{el}=\infty$ line depicts the results for technically placing a pulsed linear e-lens element next to each space charge computation node around the simulated ring. Remarkably, the purple line matches the blue $n_\mathrm{el}=24$ line, indicating both yield the same space charge limit. We conclude that the 24 e-lens configuration saturates and reaches the maximum achievable space charge limit with the suggested pulsed linear e-lens technique.

In a final step, this maximum achievable space charge limit is compared to the theoretical 2D limit of frozen longitudinal motion: here, periodic resonance crossing is suppressed not by pulsed e-lenses but by artificially freezing the longitudinal coordinates of the particles (synchrotron tune $Q_\mathrm{s}=0$) and allowing only transverse dynamics. 
The simulated intensity scans are summarized by the gray line with filled circles. The $Q_\mathrm{s}=0$ space charge limit is identified at $N=4.0\cdot N_0$, as indicated by the grey hatched area in Fig.~\ref{fig: sclimit}.
In fact, this limit is located not far beyond the maximum space charge limit with pulsed linear e-lenses at $N=2.8\cdot N_0$.
The remaining discrepancy can likely be attributed to the nonlinear part of the transverse space charge force of the ion beam, which continues to be modulated along the longitudinal plane in the pulsed linear e-lens case as opposed to the 2D $Q_\mathrm{s}=0$ case.

\begin{table}[tp]
    \caption{Space charge (SC) limit with electron lenses.}
    \label{tab: results}
    \centering
    \begin{tabular}{cccc} \hline\hline
        Number $n_\mathrm{el}$ & Compensation $\alpha$ & SC limit & SC limit [$n_\mathrm{el}=0$] \\ \hline
        0 & & $1.4\cdot N_0$ & $100\%$ \\
        3 & 0.5 & $1.8\cdot N_0$ & $130\%$ \\ 
        6 & 0.5 & $2.1\cdot N_0$ & $150\%$ \\
        12 & 0.7 & $2.6\cdot N_0$ & $185\%$ \\
        $24,\infty$ & 0.7 & $2.8\cdot N_0$ & $200\%$ \\
    \hline\hline
    \end{tabular}
\end{table}

In summary, pulsed linear electron lenses 
effectively increase the space charge limit in hadron synchrotrons, addressing the critical intensity limit 
caused by periodic resonance crossing. 
Demonstrated for the first time through full-scale simulations, 
this concept shows promise for
the example of the FAIR heavy-ion synchrotron SIS100. 
With few e-lenses, the bunch intensity limit can increase by up to $50\%$, and even double with more e-lenses.
The linear e-lens performance approaches the theoretical 2D limit of frozen longitudinal motion, where periodic resonance crossing is absent by construction.
This novel space-charge mitigation method can benefit any synchrotron with space in straight sections for a few e-lenses while respecting the lattice superperiodicity, which avoids dense systematic half-integer resonances. While exceeding typical results of other approaches, such as bunch flattening or resonance compensation, which offer $\approx 30\%$ maximum intensity increase, these techniques can complement pulsed e-lens operation to further increase the space charge limit. A prototype with a modulated \SI{10}{\ampere} electron current is underway for the SIS100 injector, SIS18. Installing three such e-lenses in SIS100 could compensate for the space charge of heavy ions like ${}^{238}$U${}^{28+}$ beyond the identified ideal $\alpha=0.5$. First experimental proof-of-principle studies with the prototype are planned at the SIS18 in the near future.

Appendix on SIS100 simulation details.---The data set containing the simulation results analyzed in this study is published in Ref.~\cite{dataset}. Figure 2 plots the instantaneous phase advance as determined from particle tracking results during a single turn in SIS100 with a symmetric configuration of six pulsed linear electron lenses, cf.\ Table \ref{tab: params}. The upper panel presents results for mono-energetic 3D Gaussian bunches, such that the lattice chromaticity has no effect on the tune footprint.

All results presented in the Letter have been based on simulations with the fixed frozen Gaussian field map approximation, which by construction only models incoherent resonances. To validate the conclusions, self-consistent particle-in-cell (PIC) simulations for the bunches have been conducted using the same detailed model as in Ref.~\cite{PhysRevAccelBeams.25.054402}), resolving the 3D bunch with \SI{10000000}{macro}-particles for the \SI{20000}{turns} cases. The tune diagram has been scanned for the $n_\mathrm{el}=6$ compensation scenario. At a bunch intensity of twice the FAIR design intensity, $N=2N_0$, the maximum space charge tune shift corresponds to $\Delta Q^\mathrm{SC}_y=-0.6$. No coherent resonances are observed outside the stopbands predicted by the frozen field map model. Large low-loss tune areas of similar size to the approximate model predictions could be identified, confirming the identified good working point areas presented in Fig.~\ref{fig: sclimit 6el}.
The identified optimum PIC-simulated working point for this scenario features about 1\% beam loss at less than $15$\% transverse rms emittance growth during \SI{20000}{turns}.

\bibliography{bibliography}

\end{document}